\documentclass[prb,showpacs,twocolumn,preprintnumbers,amsmath,groupedaddress,amssymb,superscriptaddress,floatfix]{revtex4}
\usepackage[dvips]{graphicx}
\usepackage[latin1]{inputenc}
\usepackage{xcolor,bm}
\usepackage{nicefrac}

\begin{document}
\draft

\hyphenation{a-long}

\title{Common effect of chemical and external pressures on the magnetic properties of $\bm{R}$CoPO ($\bm{R}$ = La, Pr, Nd, Sm). II.}

\author{G.~Prando}\email[E-mail: ]{g.prando@ifw-dresden.de}\affiliation{Leibniz-Institut f\"ur Festk\"orper- und Werkstoffforschung (IFW) Dresden, D-01171 Dresden, Germany}
\author{G.~Profeta}\affiliation{Department of Physical and Chemical Sciences and SPIN-CNR, Universit\`a dell'Aquila, I-67100 L'Aquila, Italy}
\author{A.~Continenza}\affiliation{Department of Physical and Chemical Sciences and SPIN-CNR, Universit\`a dell'Aquila, I-67100 L'Aquila, Italy}
\author{R.~Khasanov}\affiliation{Laboratory for Muon Spin Spectroscopy, Paul Scherrer Institut, CH-5232 Villigen PSI, Switzerland}
\author{A.~Pal}\affiliation{National Physical Laboratory (CSIR), New Delhi 110012, India}\affiliation{Department of Physics, Indian Institute of Science, Bangalore 560012, India}
\author{V.~P.~S.~Awana}\affiliation{National Physical Laboratory (CSIR), New Delhi 110012, India}
\author{B.~B\"uchner}\affiliation{Leibniz-Institut f\"ur Festk\"orper- und Werkstoffforschung (IFW) Dresden, D-01171 Dresden, Germany}\affiliation{Institut f\"ur Festk\"orperphysik, Technische Universit\"at Dresden, D-01062 Dresden, Germany}
\author{S.~Sanna}\affiliation{Dipartimento di Fisica and Unit\`a CNISM di Pavia, Universit\`a di Pavia, I-27100 Pavia, Italy}

\widetext

\begin{abstract}
The direct correspondence between Co band ferromagnetism and structural parameters is investigated in the pnictide oxides $R$CoPO for different rare-earth ions ($R$ = La, Pr, Nd, Sm) by means of muon-spin spectroscopy and {\it ab-initio} calculations, complementing our results published previously [G. Prando {\it et al.}, {\it Phys. Rev. B} {\bf 87}, 064401 (2013)]. Both the transition temperature to the ferromagnetic phase $T_{_{\textrm{C}}}$ and the volume of the crystallographic unit cell $V$ are found to be conveniently tuned by the $R$ ionic radius and/or external pressure. A linear correlation between $T_{_{\textrm{C}}}$ and $V$ is reported and {\it ab-initio} calculations unambiguously demonstrate a full equivalence of chemical and external pressures. As such, $R$ ions are shown to be influencing the ferromagnetic phase only via the induced structural shrinkage without involving any active role from the electronic $f$ degrees of freedom, which are only giving a sizeable magnetic contribution at much lower temperatures.
\end{abstract}


\date{\today}

\maketitle

\narrowtext

\section{Introduction}

In the last decade, quaternary $RMX$O oxides with tetragonal ZrCuSiAs structure (also generally referred to as $1111$ compounds) have been extensively investigated on both experimental and theoretical levels. Here, one has an alternating stacked sequence of $R$O and $MX$ layers along the $c$ axis and, within each layer, $R^{3+}$ ($X^{3-}$) ions surround O$^{2-}$ ($M^{2+}$) with a local tetrahedral crystalline arrangement. Depending on the actual choice of the pnictide element $X$ in combination with a transition metal $M$, different electronic properties can be achieved\cite{Xu08} ranging from high-$T_{_{\textrm{c}}}$ bulk superconductivity (with $T_{_{\textrm{c}}}$ values up to $55$ K for $M$ = Fe and $X$ = As)\cite{Kam08,Ren08,Joh10,Ste11,Pra11} to ferromagnetism with rare Kondo features (typically for $M$ = Ru and $X$ = P)\cite{Kre07,Kot13} or with large -- even colossal -- magnetoresistance ($M$ = Co, Mn and $X$ = As, P).\cite{Yan08,Eme10,Eme11,Oht11,Pal11a,Shi11,Wil12} However, the complex multi-orbital nature of the fermiology,\cite{Sin08,Gra09} together with the open issue about the importance of electronic correlations,\cite{Qaz09,DeM15} makes the overall understanding of these materials still not complete and currently highly debated.

The role of the rare-earth element $R$ is subtle as well. In the case of $R$FeAsO superconductors, it is well known that $R$ strongly influences the optimal-doping $T_{_{\textrm{c}}}$ value, which is more than doubled from $\sim 25$ K in the case of $R$ = La to $\sim 55$ K for $R$ = Sm. While it is excluded that this enhancement is associated with $4f$ orbitals and with the localized magnetic moment at the $R$ site,\cite{Pra10} no theory can still fully account for the experimental observations. The structural origin (i.~e., chemical pressure) seems to be a more adequate explanation for the observed effects,\cite{Pra13c} however in this case it would be naively expected that external pressure in optimally-doped systems could enhance $T_{_{\textrm{c}}}$ as well, which is not the case experimentally.\cite{Miy13,Pra15b} Chemical dilutions and quenched disorder are expected indeed to crucially drive the superconducting properties as well,\cite{Pra13b} invalidating this way a full analogy between the effects of external and chemical pressures on $T_{_{\textrm{c}}}$.\cite{Pra15b}

In this respect, the impact of structural effects on the electronic ground state of the system seems to be much easier to control in the absence of intentionally-introduced chemical disorder. This was shown to be the case for undoped $R$FeAsO materials, where external pressure ($P$) weakens the long-range ordered (LRO) antiferromagnetism (AFM) similarly to what is achieved by means of low values of electron doping.\cite{Qur10,DeR12} At the same time, we have previously shown that the Co-based ferromagnetic (FM) compounds $R$CoPO are an ideal playground to investigate the interplay of chemical and external pressures.\cite{Pra13a} There, we reported by means of muon-spin spectroscopy ($\mu^{^{+}}$SR) and density-functional theory (DFT) calculations that ferromagnetism is sizeably enhanced with increasing $P$ for both LaCoPO and PrCoPO.\cite{Pra13a} In this paper, we extend our previous investigation to the case of NdCoPO and SmCoPO. As it is well known in the literature, the magnetic moment arising from electronic $f$ shells and localized on the Nd$^{3+}$ and Sm$^{3+}$ ions plays a key role in the overall magnetic properties of the compounds\cite{Pal11b,Maj12,Maj13} similarly to the case of their $R$CoAsO analogues.\cite{Pal11a,Oht11,Oht09b,Awa10b,McG10,Oht10a,Mar10,Sug11,Pra15a} In particular, a FM-to-AFM transition is induced within the Co sublattice with decreasing temperature, followed at much lower temperatures by a magnetic LRO phase within the $R$ sublattice.\cite{Pal11b} Here, we mainly focus on the FM phase at higher temperatures by showing a strong correlation of the magnetic and structural properties. The critical temperature $T_{_{\textrm{C}}}$ proper of the FM phase is shown indeed to be linearly dependent on the volume of the unit crystallographic cell and, in turn, on the ionic radius of $R$ ions over a wide experimental range. At the same time, $T_{_{\textrm{C}}}$ linearly depends on the external pressure as well for all the investigated compounds in a quantitatively-comparable fashion. Accordingly, also supported by {\it ab-initio} calculations, we provide a detailed and unambiguous evidence for a direct correspondence between external pressure and the lattice shrinkage induced by $R$ ions. As such, these are demonstrated to affect the main ferromagnetic instability only via the induced structural shrinkage without involving any active role from the electronic $f$ degrees of freedom.

\section{Technical details}

Loose powders of LaCoPO, PrCoPO, NdCoPO and SmCoPO were grown via standard solid-state reactions\cite{Pal11b} and their structural properties were investigated at ambient temperature ($T$) and ambient $P$ by means of x-ray diffractometry (XRD). The samples crystallized in the tetragonal phase, space group $P4/nmm$, diffraction patterns and Rietveld refinements being reported in Refs.~\onlinecite{Pra13a} and \onlinecite{Pal11b}. For the aim of clarity, the lattice parameters $a$ and $c$ for all the materials are summarized in Tab.~\ref{TabLatticeConstants} together with the ionic radius value $r_{_{\textrm{I}}}$ for the corresponding rare-earth ion $R$ in a valence state $3+$ and with a coordination number $8$ (see Ref.~\onlinecite{Lid08}). A clear correlation of $r_{_{\textrm{I}}}$ with the unit cell volume $V$ is observed. It should be stressed that the chemical shrinkage of the cell driven by $R$ ions is a well-known effect for the $1111$ family of compounds.\cite{Oht09b,Luo10,Nit10}
\begin{table}[b!]
	\caption{Results of Rietveld refinements of x-ray powder diffraction patterns reported in Refs.~\onlinecite{Pra13a} and \onlinecite{Pal11b}. Lattice parameters $a$ and $c$ for the investigated compounds are reported together with the corresponding volume $V$ of the tetragonal unit cell and the ionic radius $r_{_{\textrm{I}}}$ for the $R$ ions.}
	\label{TabLatticeConstants}%
	\vspace*{0.2cm}
	\bgroup
	\begin{tabular}{ccccc}
		\hline
		\hline
		\textbf{Compound} & \textbf{$\bm{a}$ (Å)} & \textbf{$\bm{c}$ (Å)} & \textbf{$\bm{V}$ (Å$\bm{^{3}}$)} & \textbf{$\bm{r}_{_{\textrm{I}}}$ (Å)}\\
		\hline
		LaCoPO & $3.968(7)$ & $8.368(3)$ & $131.754$ & $1.18$\\
		PrCoPO & $3.921(5)$ & $8.212(4)$ & $126.253$ & $1.14$\\
		NdCoPO & $3.904(2)$ & $8.182(2)$ & $124.704$ & $1.12$\\
		SmCoPO & $3.877(4)$ & $8.072(9)$ & $121.331$ & $1.09$\\
		\hline
		\hline
	\end{tabular}
	\egroup
\end{table}

Measurements of dc magnetometry were carried out on a commercial Physical Property Measurement System (PPMS) and on a superconducting quantum interference device (SQUID) magnetometer Magnetic Property Measurement System (MPMS) by Quantum Design.

$\mu^{^{+}}$SR measurements\cite{Blu99,Yao11} were performed at the S$\mu$S muon source (Paul Scherrer Institut, Switzerland), in conditions of zero external magnetic field (ZF) and for $1.6$ K $\leq T \leq 300$ K. The GPS and Dolly spectrometers ($\pi$M3 and $\pi$E1 beamlines, respectively) were employed to perform low-background measurements at ambient pressure. These were needed as independent references for the subsequent experiments performed on the GPD spectrometer ($\mu$E1 beamline). Here, external pressures $P \leq 24$ kbar were applied at ambient $T$ by means of a double-wall piston-cylinder pressure cell (PC) made of MP$35$N alloy, the transmitting medium Daphne oil $7373$ assuring nearly-hydrostatic $P$ conditions in the experimental range.\cite{Dun10} ac susceptometry was employed in order to detect the shift of the superconducting critical temperature of a small In wire inside the PC at $T \sim 3$ K and to accurately quantify $P$ in turn. $\mu^{^{+}}$SR data analysis was performed similarly to what discussed in Ref.~\onlinecite{Pra13a}, and the same notation will be employed in the current paper for the aim of clearness. In particular, the general expression
\begin{equation}\label{EqGeneralFittingZFPCandSample}
A_{_{T}}(t) = A_{_{0}} \left[a_{_{\textrm{PC}}} \; e^{-\frac{\sigma_{_{\textrm{PC}}}^{2} t^{2}}{2} - \lambda_{_{\textrm{PC}}} t} + \left(1 - a_{_{\textrm{PC}}}\right) G_{_{T}}^{\textrm{s}}(t)\right]
\end{equation}
was employed in order to fit the time ($t$) dependence of the measured spin (de)polarization $P_{_{T}} \equiv A_{_{T}}/A_{_{0}}$ for the implanted muons ($\mu^{^{+}}$) at all the investigated $T$ values. Here, $A_{_{T}}$ is the so-called asymmetry function while $A_{_{0}}$ is an experimental spectrometer-dependent parameter accounting for the maximum value of $A$ corresponding to full spin polarization ($\sim 100$ \%). When performing experiments on the low-background GPS and Dolly spectrometers, $a_{_{\textrm{PC}}} \simeq 0$ arises from the $\mu^{^{+}}$ implanted into the sample holder, into the cryostat walls etc. At the same time, for measurements on the GPD spectrometer, $a_{_{\textrm{PC}}} \simeq 0.5$ -- $0.65$ mainly accounts for the fraction of incoming $\mu^{^{+}}$ stopping in the PC. The spin polarization is here affected by the nuclear magnetism of the MP$35$N alloy resulting in a damping governed by the Gaussian and Lorentzian parameters $\sigma_{_{\textrm{PC}}}$ and $\lambda_{_{\textrm{PC}}}$. The $T$ dependence of these quantities was independently estimated in a dedicated set of measurements on the empty PC. It should be remarked that, when ferromagnetic or superconducting samples are loaded into the PC and when performing measurements in the presence of an external magnetic field, a stray field is induced outside the sample space. This leads to a complicated modulation of the local field within the PC which can be accounted for by introducing a proper profile-function and generalizing Eq.~\eqref{EqGeneralFittingZFPCandSample} accordingly.\cite{Mai11,Pra15b}

\begin{figure*}[htbp]
	\vspace{5.4cm} \includegraphics{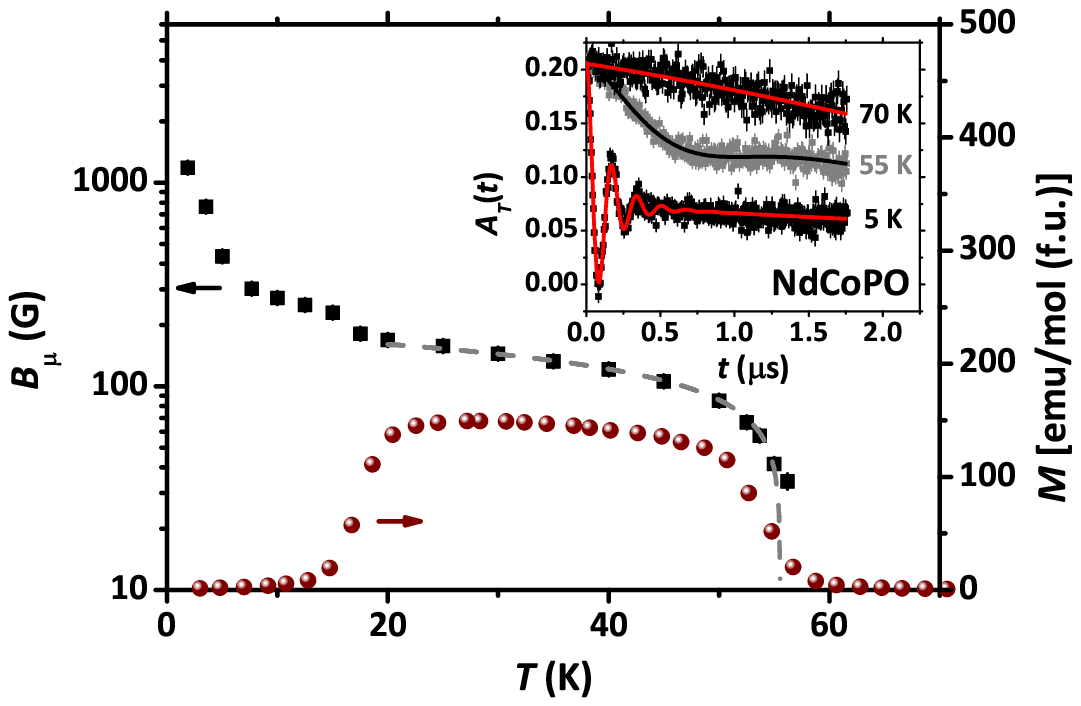} \includegraphics{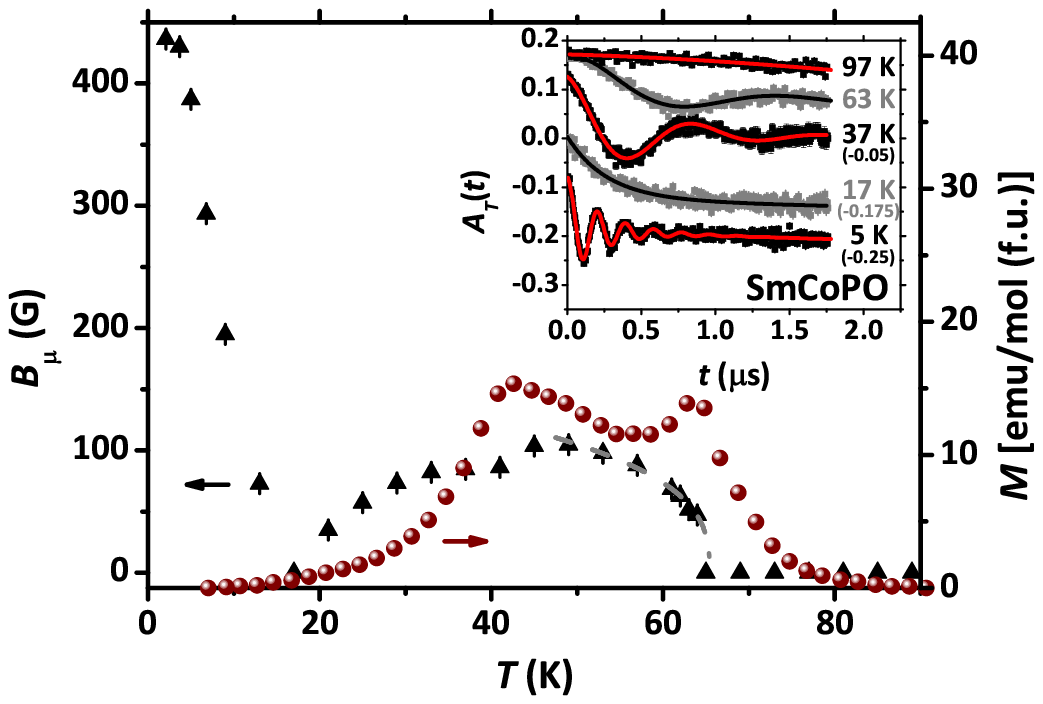}
	\caption{\label{GraInternalFieldsAmbP}(Color online) Main experimental results for NdCoPO (left panel) and SmCoPO (right-hand panel) at ambient $P$: $T$ dependence of the internal magnetic field at the $\mu^{^{+}}$ site from the calibration low-background ZF-$\mu^{^{+}}$SR measurements (black filled symbols) and of the field-cooled dc magnetization $M$ at $H = 10$ Oe (red filled circles). $M$ for SmCoPO is reported after the subtraction of a small linear term accounting for the contribution of magnetic extrinsic phases. For both samples, the dashed gray line is a best fit according to Eq.~\eqref{EqInternalFieldMeanField} with $\zeta = 0.34$ as fixed parameter. Insets: ZF-$\mu^{^{+}}$SR depolarization curves at representative $T$ values (ambient $P$, low-background spectrometers). Continuous lines are best-fitting curves according to Eqs.~\eqref{EqGeneralFittingZFPCandSample} and \eqref{EqGeneralFittingZFSample}.}
\end{figure*} 
In the case magnetic materials are being investigated, the function
\begin{eqnarray}\label{EqGeneralFittingZFSample}
G_{_{T}}^{\textrm{s}}(t) & = &
\left[1 - V_{_{\textrm{m}}}(T)\right] e^{-\frac{\sigma_{_{\textrm{N}}}^{2} t^{2}}{2}} +{}\nonumber\\ & + & \left[a^{\perp}(T) F(t) D^{\perp}(t) + a^{\parallel}(T) D^{\parallel}(t)\right].
\end{eqnarray}
is employed in order to describe the spin (de)polarization arising from the remaining fraction $\left(1 - a_{_{\textrm{PC}}}\right)$ of $\mu^{^{+}}$, that is, the particles implanted into the sample. $V_{_{\textrm{m}}}(T)$ is the fraction of $\mu^{^{+}}$ probing static local magnetic fields (i.~e., the magnetic volume fraction of the investigated material, due to the random implantation of $\mu^{^{+}}$). For $V_{_{\textrm{m}}}(T) = 0$, only nuclear magnetic moments can cause a damping of the signal. Typically this signal is of Gaussian nature with characteristic rate $\sigma_{_{\textrm{N}}} \sim 0.1 \; \mu$s$^{-1}$. Below the critical transition temperature $T_{_{\textrm{C}}}$ to the ferromagnetic phase, the superscript $\perp$ ($\parallel$) refers to $\mu^{^{+}}$ probing a local static magnetic field in a perpendicular (parallel) direction with respect to the initial $\mu^{^{+}}$ spin polarization and, accordingly, one has $\left[a^{\perp}(T) + a^{\parallel}(T)\right] = V_{_{\textrm{m}}}(T)$ for the so-called ``transversal'' ($a^{\perp}$) and ``longitudinal'' ($a^{\parallel}$) fractions. A coherent precession of $\mu^{^{+}}$ around the local field $B_{_{\mu}}$ generated by a LRO magnetic order can be observed in the transversal amplitude. This is typically described by the oscillating function $F(t)$, the standard choice $F(t) = \cos\left(\gamma B_{_{\mu}} t + \phi\right)$ being in good agreement with current experimental data (statistical $\chi^{2} \simeq 1 - 1.2$) at all $T$ values. Here, $\gamma = 2 \pi \times 135.54$ MHz/T is the gyromagnetic ratio for $\mu^{^{+}}$, while typical phase $\left|\phi\right| \sim 20$° -- $30$° are measured (experimentally, negative values are found -- see also the discussion in Ref.~\onlinecite{Pra13a}). The damping function $D^{\perp}(t)$ reflects a static distribution of local magnetic fields while the longitudinal component typically probes dynamical spin-lattice-like relaxation processes resulting in an exponential damping function $D^{\parallel}(t)$.

First principles calculations were performed using the VASP package\cite{Kre96a,Kre96b} within the generalized gradient approximation (GGA) to DFT\cite{Per96} and projected augmented-wave (PAW)\cite{Blo94} pseudopotentials for both the non-magnetic and ferromagnetic structures. Full relaxation of the cell shape and of the internal positions was achieved using $650$ eV as energy cutoff and a ($12$, $12$, $6$) $k$-point mesh within the Monkhorst-Pack scheme\cite{Mon76} and making the {\it ab-initio} forces to vanish (up to  $0.001$ eV/Å) at each fixed cell volume.

\section{Experimental results}

\subsection{Summary of main results from Ref.~\onlinecite{Pra13a} on LaCoPO and PrCoPO}

In a previous work,\cite{Pra13a} we reported on a $\mu^{^{+}}$SR investigation of magnetism in LaCoPO and PrCoPO. A LRO-FM phase with critical temperature $T_{_{\textrm{C}}} = 33.2 \pm 1.0$ K and $T_{_{\textrm{C}}} = 48.0 \pm 1.0$ K develops for LaCoPO and PrCoPO, respectively. The $T$ dependence of the local magnetic field at the $\mu^{^{+}}$ site $B_{\mu}$ is well described for $T \leq T_{_{\textrm{C}}}$ by a power-law function\cite{Pra13a}
\begin{equation}\label{EqInternalFieldMeanField}
B_{_{\mu}}(T) = B_{_{\mu}}(0)\left(1-\frac{T}{T_{_{\textrm{C}}}}\right)^{\zeta}
\end{equation}
where $\zeta = 0.34 \pm 0.01$ for both LaCoPO and PrCoPO. In spite of the qualitatively identical behaviour, La$^{3+}$ is a non-magnetic ion at variance with Pr$^{3+}$. Accordingly, the only contribution to $B_{_{\mu}}(T)$ must arise from the Co sublattice alone in both the cases and, moreover, the functional form in Eq.~\eqref{EqInternalFieldMeanField} must be intrinsically associated to the FM phase from Co. The observed behaviour is \textit{qualitatively} preserved independently on the $P$ value, while a clear dependence on $P$ was detected for both $B_{_{\mu}}(0)$ and $T_{_{\textrm{C}}}$. A detailed discussion was presented in Ref.~\onlinecite{Pra13a} concerning the former quantity. Here, we want to concentrate on $T_{_{\textrm{C}}}(P)$, which was found to linearly increase with $P$ in both LaCoPO an PrCoPO.\cite{Pra13a}
\begin{figure*}[htbp]
	\vspace{5.6cm} \includegraphics{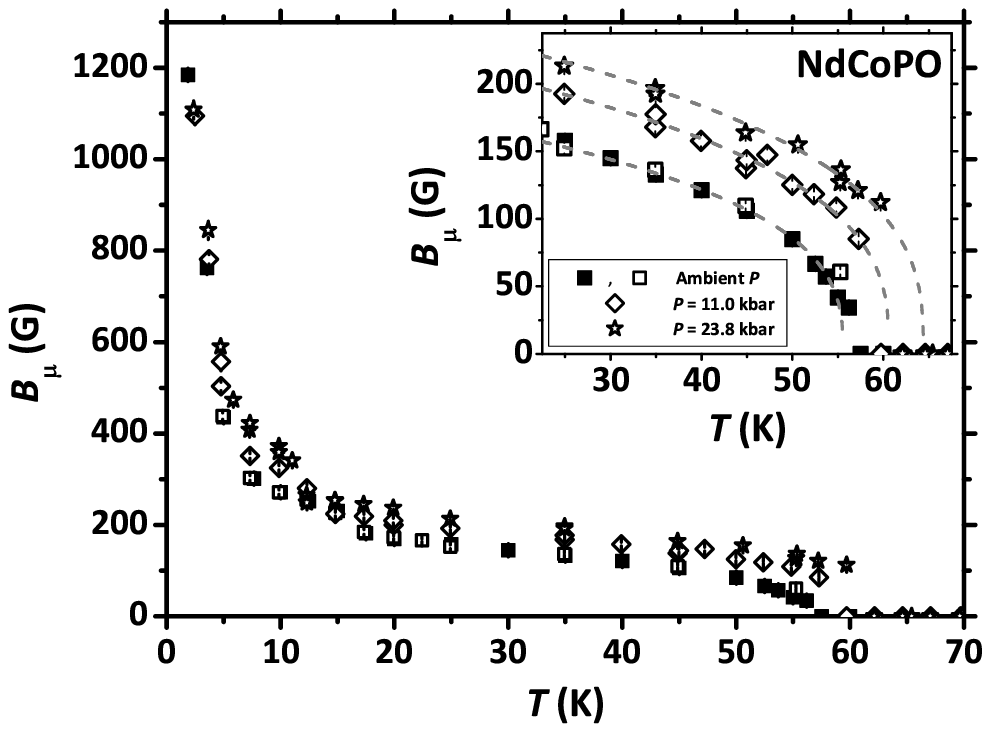} \includegraphics{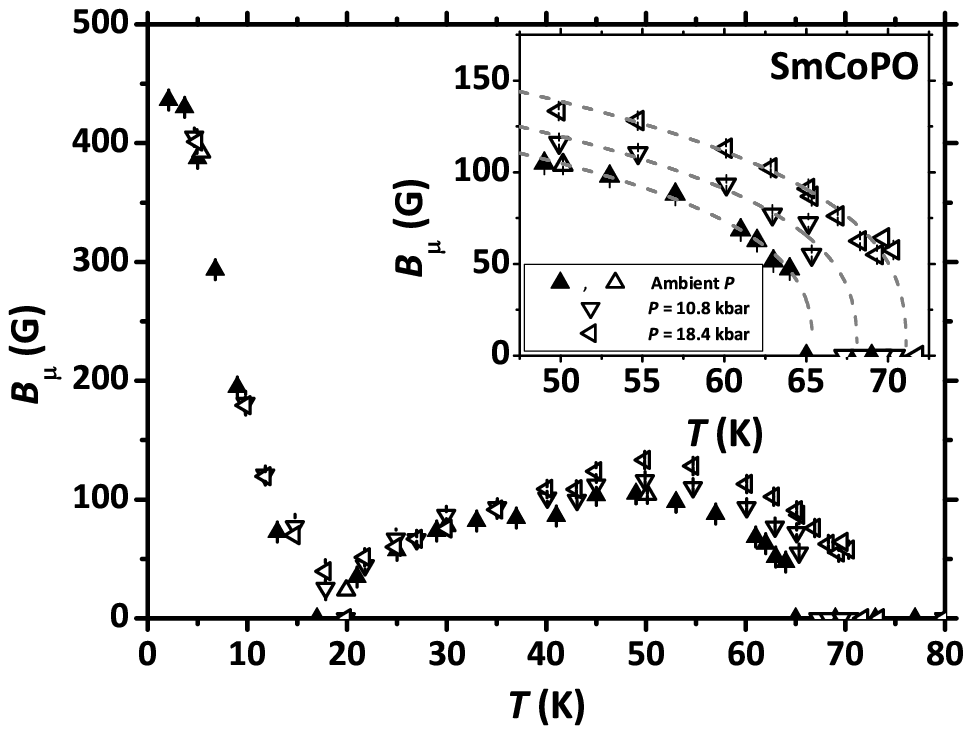}
	\caption{\label{GraInternalFieldsUnderP}$T$ dependence of the internal field at the $\mu^{^{+}}$ site for NdCoPO (left panel) and SmCoPO (right-hand panel) for $P < 24$ kbar (open symbols). Black filled symbols are relative to low-background measurements (reproduced from Fig.~\ref{GraInternalFieldsAmbP}). Insets display an enlargement of the $T$ region just below $T_{_{\textrm{C}}}$. For both samples, the dashed gray lines are best fits according to Eq.~\eqref{EqInternalFieldMeanField} with $\zeta = 0.34$ as fixed parameter.}
\end{figure*}

\subsection{NdCoPO and SmCoPO at ambient pressure}

The current $\mu^{^{+}}$SR measurements on NdCoPO and SmCoPO clearly evidence a different scenario if compared to what was reported for LaCoPO and PrCoPO. In fact, the magnetic moments of Nd$^{3+}$ and Sm$^{3+}$ ions crucially dominate the magnetic properties at low $T$ and the interplay of $d$ and $f$ electrons leads to progressive re-ordering effects on the Co sublattice.\cite{Pal11b,Maj13} This is confirmed by our measurements of ZF-$\mu^{^{+}}$SR and dc magnetometry both for NdCoPO and SmCoPO at ambient $P$.

As shown in the insets of Fig.~\ref{GraInternalFieldsAmbP}, well-defined spontaneous coherent oscillations appear in the $\mu^{^{+}}$SR depolarization curves below $\sim 55$ K and $\sim 65$ K for NdCoPO and SmCoPO, respectively, indicating the development of a magnetic LRO phase in both the samples. These values also correspond to the sudden increase of the dc magnetization $M$, as reported in the main panels of Fig.~\ref{GraInternalFieldsAmbP}. Just below this transition, $B_{_{\mu}}(T)$ can be well reproduced by the power-law expression reported in Eq.~\eqref{EqInternalFieldMeanField}, again with $\zeta \simeq 0.34$. By assuming that $\zeta = 0.34$ (fixed parameter), a fitting of $B_{_{\mu}}(T)$ data leads to the estimates $T_{_{\textrm{C}}} = 55.5 \pm 1.0$ K and $T_{_{\textrm{C}}} = 65.4 \pm 1.0$ K for NdCoPO and SmCoPO, respectively (see the dashed lines in the main panels of Fig.~\ref{GraInternalFieldsAmbP}). However, such trend for $B_{_{\mu}}(T)$ is no longer obeyed below $T_{_{\textrm{N}}} \sim 16$ K and $T_{_{\textrm{N}}} \sim 45$ K for NdCoPO and SmCoPO, respectively, where clear anomalies in the $M(T)$ curves are also detected. With further decreasing $T$ below $T_{_{\textrm{N}}}$, $M$ vanishes while $B_{_{\mu}}(T)$ diverges to positive values in the case of NdCoPO, while a negative divergence is rather observed for SmCoPO [it should be recalled that the absolute value of $B_{_{\mu}}(T)$ is experimentally accessed indeed].

These results can be understood as follows. As $B_{_{\mu}}(T)$ is well reproduced by Eq.~\eqref{EqInternalFieldMeanField} with $\zeta = 0.34$ for $T_{_{\textrm{N}}} < T < T_{_{\textrm{C}}}$, it is reasonable to associate the magnetic transition at $T_{_{\textrm{C}}}$ with a FM phase originating from Co similarly to what is achieved in LaCoPO and PrCoPO. In correspondence to $T_{_{\textrm{N}}}$ another transition on the Co sublattice develops with globally AFM correlations, as indicated by the gradual vanishing of $M$, possibly favoured by the magnetic moments of the $R$ ions. The positive and negative divergences of $B_{_{\mu}}(T)$ in NdCoPO and SmCoPO, respectively, may be originating from several reasons among which, e.~g., different microscopic magnetic configurations or statistical occupancies of the two $\mu^{^{+}}$ crystallographic sites in the two materials.\cite{Pra15a,Bon15} These issues are beyond the scopes of this paper and we will concentrate only on $T_{_{\textrm{C}}}$ from now on.

\subsection{NdCoPO and SmCoPO for non-zero external pressure}

The impact of the crystallographic structure on magnetism in NdCoPO and SmCoPO has been further investigated through the application of an external pressure. As main result (see Fig.~\ref{GraInternalFieldsUnderP}), no qualitative changes are induced by $P$ in the $T$ dependence of $B_{_{\mu}}$ whose shape is preserved identical to what is reported in Fig.~\ref{GraInternalFieldsAmbP}. However, on a quantitative level, $T_{_{\textrm{C}}}$ is enhanced by $P$ in a linear fashion for both NdCoPO and SmCoPO, as enlightened in the insets of Fig.~\ref{GraInternalFieldsUnderP}. Overall, the situation is extremely reminiscent of what was observed in LaCoPO and PrCoPO, where $P$ induces a linear increase of $T_{_{\textrm{C}}}$ as well.\cite{Pra13a} Similarly to those compounds and as shown in the insets of Fig.~\ref{GraInternalFieldsUnderP}, the shape of $B_{_{\mu}}(T)$ is not affected at all by $P$ just below $T_{_{\textrm{C}}}$. Accordingly, Eq.~\eqref{EqInternalFieldMeanField} can still be used for $T > T_{_{\textrm{N}}}$ as a fitting function with fixed parameter $\zeta = 0.34$ in order to estimate $T_{_{\textrm{C}}}(P)$ (see Fig.~\ref{GraTcVsP} later on). At the same time, $T_{_{\textrm{N}}}$ is clearly suppressed upon increasing $P$ for NdCoPO. In particular, $T_{_{\textrm{N}}} = \left(16.25 \pm 1.25\right)$ K at ambient $P$ shifts to $T_{_{\textrm{N}}} = \left(11.7 \pm 0.7\right)$ K at $P = 23.8$ kbar. The simultaneous $T_{_{\textrm{C}}}$ enhancement and $T_{_{\textrm{N}}}$ suppression is reminiscent of the effect of increasing magnetic field both in NdCoPO\cite{Pal11b} and in NdCoAsO.\cite{Oht11} On the other hand $T_{_{\textrm{N}}}$ seems to be mostly unaffected by $P$ for SmCoPO -- however, in this case the density of experimental points does not allow to make any definite statement. We will no longer come back to the issue of the $T_{_{\textrm{N}}}(P)$ dependence and only focus on $T_{_{\textrm{C}}}(P)$ from now on.

\section{Discussion}

\begin{figure}[b!]
	\vspace{6.2cm} \includegraphics{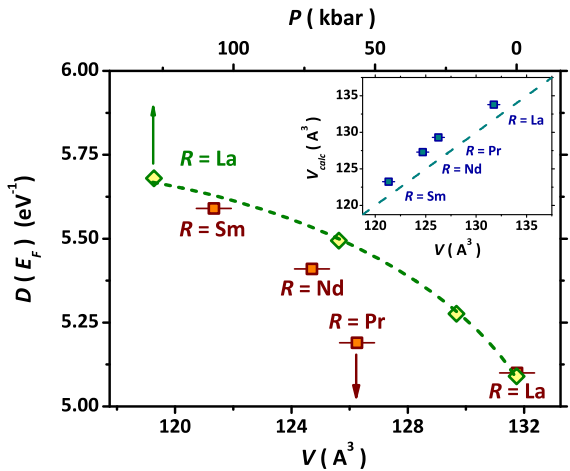}
	\caption{\label{theory}(Color online) Main panel: density of states at the Fermi level $D\left(E_{_{F}}\right)$ for the investigated $R$CoPO compounds as a function of the experimental value of the unit cell volume $V$ (open squares, relative to lower x-axis). The $D\left(E_{_{F}}\right)$ values calculated for the LaCoPO compound are also reported at different volumes corresponding to pressure values indicated on the upper x-axis (open diamonds). Inset: dependence of the calculated unit cell volume in the FM state $V_{_{calc}}$ on the experimental $V$ for the four investigated $R$CoPO compounds. The dashed line represents the condition $V_{_{calc}} = V$.}
\end{figure}
For itinerant ferromagnets, the magnetic instability is generally considered to be marked by the Stoner factor $S = I_{_{xc}} \cdot D(E_{_{F}})$ becoming larger than $1$. Here, $I_{_{xc}}$ is the exchange integral, measuring the strength of the exchange interaction sensitive to the localization of the wave functions, while $D(E_{_{F}})$ represents the density of states at the Fermi level, whose main contributions come from Co $d$-like components. An estimate of $D(E_{_{F}})$ can be achieved on an \textit{ab-initio} grounds at the experimental lattice constants of $R$CoPO at ambient pressure, as displayed by the red squares in the main panel of Fig.~\ref{theory}. Accordingly, $S$ can be computed as well by taking a value $I_{_{xc}} = 0.99$ for cobalt.\cite{Jan77} As a result, we find that the Stoner condition for FM instability is always safely satisfied for all the considered $R$CoPO compounds. Indeed, the FM state has a lower energy than the non-magnetic one, the total energy difference being of the order of $50-60$ meV/cell for every compound. The green diamonds in the main panel of Fig.~\ref{theory} are $D(E_{_{F}})$ values calculated for LaCoPO under external pressure and, remarkably, these values very closely mimic the trend of the different compounds at their respective equilibrium volumes, showing that the main variation is essentially linked to the internal pressure effect induced by the smaller $R$. This provides evidence that the change in the electronic properties of $R$CoPO is merely due to structure and that the $f$ states associated with the $R$ ions do play a likely passive role. As shown in the inset of Fig.~\ref{theory}, the calculated equilibrium volumes $V_{_{calc}}$ in the ferromagnetic state for all the considered compounds reproduce very nicely the experimental trend of $V$ as a function of $R$ (see the dashed line in the inset) both qualitatively and quantitatively within the usual DFT-GGA accuracy. The calculated magnetic moment of Co atoms decreases with reducing the volume of the tetragonal cell from $0.52$ $\mu_{_{\textrm{B}}}$ in LaCoPO to $0.48$ $\mu_{_{\textrm{B}}}$ for SmCoPO (not shown). While this trend is in qualitative agreement with the experimental results for LaCoPO and PrCoPO,\cite{Pra13a} the strong magnetic contribution from $R$ ions makes an experimental check by means of dc magnetometry uncertain for both NdCoPO and SmCoPO.

\begin{figure}[t!]
	\vspace{5.8cm} \includegraphics{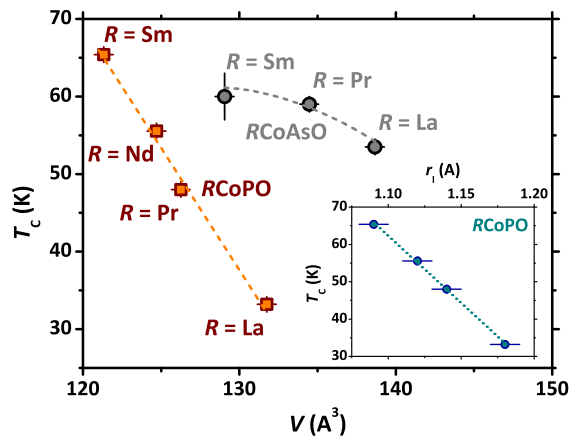}
	\caption{\label{GraCompAsVsPPlusIonicRadius}(Color online) Main panel: dependence of the critical temperature $T_{_{\textrm{C}}}$ on the measured volume $V$ of the tetragonal unit cell at ambient $P$ for $R$CoPO and for the isostructural $R$CoAsO (these latter data are taken from Ref.~\onlinecite{Pra15a}). The orange dashed curve is a linear best fitting to $R$CoPO data while the gray curve is a guide to the eye. Inset: dependence of the critical temperature $T_{_{\textrm{C}}}$ on the ionic radius $r_{_{\textrm{I}}}$ of $R$ ions -- see Tab.~\ref{TabLatticeConstants}. The dotted curve is a linear best-fittings to data.}
\end{figure}
We have shown above that, due to the effect of magnetic $R$ ions, the actual $T$ dependence of $B_{_{\mu}}(T)$ for NdCoPO and SmCoPO strongly departs below $T_{_{\textrm{N}}}$ from the behaviour reported for LaCoPO and PrCoPO. However, just below $T_{_{\textrm{C}}}$, strong qualitative analogies among all the samples suggest a common origin of the FM phase. To make these arguments more quantitative, we report in the main panel of Fig.~\ref{GraCompAsVsPPlusIonicRadius} the $T_{_{\textrm{C}}}$ values for the four $R$CoPO samples as a function of the relative measured $V$ values for the tetragonal unit cell at ambient $T$ and $P$ and, in the inset, as a function of the $r_{_{\textrm{I}}}$ of the corresponding $R$ ion (see Tab.~\ref{TabLatticeConstants}). Not only it is evident that the lattice shrinkage triggered by smaller ionic radii is inducing a dramatic increase of $T_{_{\textrm{C}}}$ (being even doubled by a full replacement of La with Sm), but a linear correlation is clearly observed as well between $T_{_{\textrm{C}}}$ and both $V$ and $r_{_{\textrm{I}}}$.

According to some theoretical models,\cite{Moh87,Kai88} $D(E_{_{F}})$ is expected to be directly related to $T_{_{\textrm{C}}}$ and, as discussed above, we found that it slightly increases with the atomic radius of $R$ and/or with external pressure. Although the increase of $D(E_{_{F}})$ will surely result in a positive contribution to $T_{_{\textrm{C}}}$, we find from calculations that the small actual variation in its absolute value with chemical or external pressures can not explain the large differences reported in Fig.~\ref{GraCompAsVsPPlusIonicRadius} for $T_{_{\textrm{C}}}$. Moreover, it should be remarked that the peculiar linear relation between structural parameters and magnetism is not at all a general feature of Co in $1111$ oxypnictides. In particular, the main panel of Fig.~\ref{GraCompAsVsPPlusIonicRadius} also reports $T_{_{\textrm{C}}}$ vs. $V$ data for the isostructural compounds $R$CoAsO.\cite{Pra15a} Here, the linear relation among these two quantities ceases as if the internal pressure effects are saturated yielding to a constant $T_{_{\textrm{C}}}$ when approaching the smaller size of Sm. Accordingly, no change of $T_{_{\textrm{C}}}$ has been observed as a function of external pressure as well in SmCoAsO.\cite{Pra15a} The differences between $R$CoAsO and $R$CoPO compounds imply a non-trivial difference in the microscopic origin of the link between structure and magnetism. This may likely involve different degrees of hybridization between the pnictogen and the transition metal and, in turn, quantitatively different exchange paths.
\begin{figure}[t!]
	\vspace{6.0cm} \includegraphics{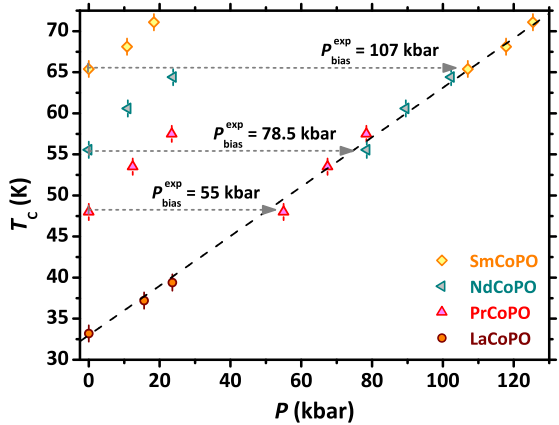}
	\caption{\label{GraTcVsP}(Color online) Summarizing results of $T_{_{\textrm{C}}}$ vs. $P$ for the four investigated $R$CoPO samples. The slope of the dashed curve is obtained by Eq.~\eqref{EqCompress} -- see text. Data for PrCoPO, NdCoPO and SmCoPO have been shifted horizontally by the sample-specific amounts $P_{_{\textrm{bias}}}^{^{\textrm{exp}}}$ as discussed in the text.}
\end{figure}

\begin{table}[b!]
	\caption{Experimental $\left(P_{_{\textrm{bias}}}^{^{\textrm{exp}}}\right)$ and computed $\left(P_{_{\textrm{bias}}}^{^{\textrm{theo}}}\right)$ values of the bias external pressures mimicking the chemical pressure from $R$ ions for all the considered compounds (see text).}
	\label{TabBias}%
	\vspace*{0.2cm}
	\bgroup
	\begin{tabular}{ccc}
		\hline
		\hline
		\textbf{Compound} & $P_{_{\textrm{bias}}}^{^{\textrm{exp}}}$ (kbar) & $P_{_{\textrm{bias}}}^{^{\textrm{theo}}}$ (kbar) \vspace{0.1cm}\\
		\hline
		LaCoPO & -- & --\\
		PrCoPO & $55$ & $56.4$\\
		NdCoPO & $78.5$ & $72.3$\\
		SmCoPO & $107$ & $106.9$\\
		\hline
		\hline
	\end{tabular}
	\egroup
\end{table}
Experimental results for the actual dependence of $T_{_{\textrm{C}}}$ on $P$ in $R$CoPO are summarized in Fig.~\ref{GraTcVsP}. It is apparent that pressure leads to a linear enhancement of $T_{_{\textrm{C}}}$ with a quantitatively similar slope for all the considered $R$CoPO compounds. By considering the observations above concerning the linear dependence of $T_{_{\textrm{C}}}$ on $V$ at ambient pressure, it is possible to discuss the close correspondence between chemical and external pressures more in detail. In particular, it is straightforward to speculate that external pressure linearly influences $V$ which, in turn, linearly enhances $T_{_{\textrm{C}}}$. These arguments can be substantiated by computing the isothermal lattice compressibility $\beta = - \left(1/V\right) \times \left(d V /d P\right)$ for LaCoPO by means of first-principles calculations. The obtained values are $7.52 \times 10^{-4} \; \textrm{kbar}^{-1}$ and $7.23 \times 10^{-4} \; \textrm{kbar}^{-1}$ in the paramagnetic and FM states, respectively. The assumption about the linear influence of both chemical and external pressures on $V$ can now be exploited by writing
\begin{equation}\label{EqCompress}
\beta = - \frac{1}{V} \frac{d V}{d P} = - \frac{1}{V} \frac{\partial V}{\partial T_{_{\textrm{C}}}} \frac{\partial T_{_{\textrm{C}}}}{\partial P}
\end{equation}
and, accordingly, an estimate can be given for the partial derivative $\partial T_{_{\textrm{C}}} / \partial P$ starting from the values computed above (the normalization factor $1/V$ is obtained from the ambient $P$ value $V = 131.754$ Å$^{3}$). This method is necessarily approximate, as $V$ values at room $T$ are involved together with $T_{_{\textrm{C}}}$, whose values are $\sim 1$ order of magnitude smaller. However, this uncertainty can be approximately solved in Eq.~\eqref{EqCompress} by selecting an intermediate value for the computed compressibility, namely $\beta \simeq 7.4 \times 10^{-4} \; \textrm{kbar}^{-1}$. This is admissible also in view of the tiny discrepancy ($\sim 4$ \%) among the two $\beta$ values reported above. A line with slope given by the calculated $\partial T_{_{\textrm{C}}} / \partial P$ is reported in Fig.~\ref{GraTcVsP} as well. One observes that experimental data are well-fitted by this line if the effect of chemical pressure introduced by $R$ ions is mimicked by constant $P_{_{\textrm{bias}}}$ offset values characteristic of each sample. The experimental values $P_{_{\textrm{bias}}}^{^{\textrm{exp}}}$ can be obtained from Fig.~\ref{GraTcVsP} by maximizing the agreement between experimental data and the line. This linear correlation strongly supports the picture that $R$ substitution acts as an effective external isotropic pressure. A more stringent and physical argument to support this picture comes from the calculation of $P_{bias}$ from first principles, with $P_{_{\textrm{bias}}}^{^{\textrm{theo}}}$ defined as the calculated pressure needed to make LaCoPO match exactly the $R$CoPO equilibrium volumes. As shown in Tab.~\ref{TabBias}, the excellent agreement between $P_{_{\textrm{bias}}}^{^{\textrm{exp}}}$ and $P_{_{\textrm{bias}}}^{^{\textrm{theo}}}$ for all the considered samples is a clear indication towards the full equivalence of chemical and external pressures in $R$CoPO.

\section{Conclusions}

In this paper, we have extended our previous $\mu^{+}$SR investigation of itinerant ferromagnetism in LaCoPO and PrCoPO to the cases of NdCoPO and SmCoPO. The magnetic moment arising from electronic $f$ shells and localized on the Nd$^{3+}$ and Sm$^{3+}$ ions plays a key role in the overall magnetic properties of the compounds at low temperatures. However, as long as only the ferromagnetic phase at higher temperatures is considered, a strong correlation of the magnetic and structural properties is reported and discussed. The critical temperature $T_{_{\textrm{C}}}$ proper of the ferromagnetic phase is linearly dependent on the volume of the unit crystallographic cell and on the ionic radius of $R$ ions over a wide experimental range. Its value can be even doubled by a full replacement of La with Sm. At the same time, $T_{_{\textrm{C}}}$ linearly depends on the external pressure as well for all the investigated compounds in a quantitatively-comparable fashion. Accordingly, also supported by {\it ab-initio} calculations, we provide a detailed and unambiguous evidence for a direct correspondence between the structural shrinkage induced by external and chemical pressures. As such, the ferromagnetic instability is driven in these compounds only by the crystallographic structure without involving any active role from the electronic $f$ degrees of freedom. Calculations based on the variation in the density of states at the Fermi energy upon increasing pressures are not able to quantitatively reproduce the full enhancement $T_{_{\textrm{C}}}$, opening the way to further theoretical investigations of these materials. In this respect, we recall that 2D--3D crossover for spin-fluctuations have been put forward to explain previous experimental results\cite{Oht09b,Maj13} and that a first-principle finite-temperature description of itinerant ferromagnets remains a big challenge in condensed matter theory.

\section{Acknowledgements}

We acknowledge useful discussions with C. Ortix. G. Prando acknowledges support by the Humboldt Research Fellowship for Postdoctoral researchers. S. Sanna acknowledges partial support of PRIN2012 Project No. 2012X3YFZ2. A. Pal acknowledges financial support from Dr. D. S. Kothari Postdoctoral Fellowship (UGC-DSKPDF) Program. This research at NPL, New Delhi, is supported by the DAE-SRC outstanding Investigator Award Scheme. The experimental $\mu^{^{+}}$SR work was performed at the Swiss Muon Source (S$\mu$S) at the Paul Scherrer Institut, Switzerland.



\end{document}